\documentclass[conference]{worldcomp}

\newtheorem{theorem}{Theorem}
\usepackage{graphicx}
\newdimen\proofrulebreadth \proofrulebreadth=.05em
\newdimen\proofdotseparation \proofdotseparation=1.25ex
\newdimen\proofrulebaseline \proofrulebaseline=2ex
\newcount\proofdotnumber \proofdotnumber=3
\let\then\relax
\def\hfi{\hskip0pt plus.0001fil}
\mathchardef\squigto="3A3B
\newif\ifinsideprooftree\insideprooftreefalse
\newif\ifonleftofproofrule\onleftofproofrulefalse
\newif\ifproofdots\proofdotsfalse
\newif\ifdoubleproof\doubleprooffalse
\let\wereinproofbit\relax
\newdimen\shortenproofleft
\newdimen\shortenproofright
\newdimen\proofbelowshift
\newbox\proofabove
\newbox\proofbelow
\newbox\proofrulename
\def\shiftproofbelow{\let\next\relax\afterassignment\setshiftproofbelow\dimen0 }
\def\shiftproofbelowneg{\def\next{\multiply\dimen0 by-1 }%
\afterassignment\setshiftproofbelow\dimen0 }
\def\setshiftproofbelow{\next\proofbelowshift=\dimen0 }
\def\setproofrulebreadth{\proofrulebreadth}

\def\prooftree{
\ifnum  \lastpenalty=1
\then   \unpenalty
\else   \onleftofproofrulefalse
\fi
\ifonleftofproofrule
\else   \ifinsideprooftree
        \then   \hskip.5em plus1fil
        \fi
\fi
\bgroup
\setbox\proofbelow=\hbox{}\setbox\proofrulename=\hbox{}%
\let\justifies\proofover\let\leadsto\proofoverdots\let\Justifies\proofoverdbl
\let\using\proofusing\let\[\prooftree
\ifinsideprooftree\let\]\endprooftree\fi
\proofdotsfalse\doubleprooffalse
\let\thickness\setproofrulebreadth
\let\shiftright\shiftproofbelow \let\shift\shiftproofbelow
\let\shiftleft\shiftproofbelowneg
\let\ifwasinsideprooftree\ifinsideprooftree
\insideprooftreetrue
\setbox\proofabove=\hbox\bgroup$\displaystyle 
\let\wereinproofbit\prooftree
\shortenproofleft=0pt \shortenproofright=0pt \proofbelowshift=0pt
\onleftofproofruletrue\penalty1
}

\def\eproofbit{
\ifx    \wereinproofbit\prooftree
\then   \ifcase \lastpenalty
        \then   \shortenproofright=0pt  
        \or     \unpenalty\hfil         
        \or     \unpenalty\unskip       
        \else   \shortenproofright=0pt  
        \fi
\fi
\global\dimen0=\shortenproofleft
\global\dimen1=\shortenproofright
\global\dimen2=\proofrulebreadth
\global\dimen3=\proofbelowshift
\global\dimen4=\proofdotseparation
\global\count255=\proofdotnumber
$\egroup  
\shortenproofleft=\dimen0
\shortenproofright=\dimen1
\proofrulebreadth=\dimen2
\proofbelowshift=\dimen3
\proofdotseparation=\dimen4
\proofdotnumber=\count255
}

\def\proofover{
\eproofbit 
\setbox\proofbelow=\hbox\bgroup 
\let\wereinproofbit\proofover
$\displaystyle
}%
\def\proofoverdbl{
\eproofbit 
\doubleprooftrue
\setbox\proofbelow=\hbox\bgroup 
\let\wereinproofbit\proofoverdbl
$\displaystyle
}%
\def\proofoverdots{
\eproofbit 
\proofdotstrue
\setbox\proofbelow=\hbox\bgroup 
\let\wereinproofbit\proofoverdots
$\displaystyle
}%
\def\proofusing{
\eproofbit 
\setbox\proofrulename=\hbox\bgroup 
\let\wereinproofbit\proofusing
\kern0.3em$
}

\def\endprooftree{
\eproofbit 
  \dimen5 =0pt
\dimen0=\wd\proofabove \advance\dimen0-\shortenproofleft
\advance\dimen0-\shortenproofright
\dimen1=.5\dimen0 \advance\dimen1-.5\wd\proofbelow
\dimen4=\dimen1
\advance\dimen1\proofbelowshift \advance\dimen4-\proofbelowshift
\ifdim  \dimen1<0pt
\then   \advance\shortenproofleft\dimen1
        \advance\dimen0-\dimen1
        \dimen1=0pt
        \ifdim  \shortenproofleft<0pt
        \then   \setbox\proofabove=\hbox{%
                        \kern-\shortenproofleft\unhbox\proofabove}%
                \shortenproofleft=0pt
        \fi
\fi
\ifdim  \dimen4<0pt
\then   \advance\shortenproofright\dimen4
        \advance\dimen0-\dimen4
        \dimen4=0pt
\fi
\ifdim  \shortenproofright<\wd\proofrulename
\then   \shortenproofright=\wd\proofrulename
\fi
\dimen2=\shortenproofleft \advance\dimen2 by\dimen1
\dimen3=\shortenproofright\advance\dimen3 by\dimen4
\ifproofdots
\then
        \dimen6=\shortenproofleft \advance\dimen6 .5\dimen0
        \setbox1=\vbox to\proofdotseparation{\vss\hbox{$\cdot$}\vss}%
        \setbox0=\hbox{%
                \advance\dimen6-.5\wd1
                \kern\dimen6
                $\vcenter to\proofdotnumber\proofdotseparation
                        {\leaders\box1\vfill}$%
                \unhbox\proofrulename}%
\else   \dimen6=\fontdimen22\the\textfont2 
        \dimen7=\dimen6
        \advance\dimen6by.5\proofrulebreadth
        \advance\dimen7by-.5\proofrulebreadth
        \setbox0=\hbox{%
                \kern\shortenproofleft
                \ifdoubleproof
                \then   \hbox to\dimen0{%
                        $\mathsurround0pt\mathord=\mkern-6mu%
                        \cleaders\hbox{$\mkern-2mu=\mkern-2mu$}\hfill
                        \mkern-6mu\mathord=$}%
                \else   \vrule height\dimen6 depth-\dimen7 width\dimen0
                \fi
                \unhbox\proofrulename}%
        \ht0=\dimen6 \dp0=-\dimen7
\fi
\let\doll\relax
\ifwasinsideprooftree
\then   \let\VBOX\vbox
\else   \ifmmode\else$\let\doll=$\fi
        \let\VBOX\vcenter
\fi
\VBOX   {\baselineskip\proofrulebaseline \lineskip.2ex
        \expandafter\lineskiplimit\ifproofdots0ex\else-0.6ex\fi
        \hbox   spread\dimen5   {\hfi\unhbox\proofabove\hfi}%
        \hbox{\box0}%
        \hbox   {\kern\dimen2 \box\proofbelow}}\doll%
\global\dimen2=\dimen2
\global\dimen3=\dimen3
\egroup 
\ifonleftofproofrule
\then   \shortenproofleft=\dimen2
\fi
\shortenproofright=\dimen3
\onleftofproofrulefalse
\ifinsideprooftree
\then   \hskip.5em plus 1fil \penalty2
\fi
}

\usepackage[hmargin=.75in,vmargin=1in]{geometry}
\usepackage[american]{babel}
\usepackage[T1]{fontenc}
\usepackage{times}
\usepackage{caption}

\usepackage{textcomp}
\usepackage{epsfig,graphicx}
\usepackage{xcolor}
\usepackage{amsfonts,amsmath,amssymb}
\usepackage{fixltx2e} 
\usepackage{booktabs}

\title{\bf Probabilistic Alias Analysis for Parallel Programming in SSA Forms}           

\author{
{\bfseries Mohamed A. El-Zawawy$^1$ and Mohammad N. Alanazi$^2$}\\
$^{1,2}$College of Computer and Information Sciences,\\ Al Imam
Mohammad Ibn Saud
Islamic University (IMSIU)\\ Riyadh, Kingdom of Saudi Arabia\\
$^1$Department of Mathematics, Faculty of Science
\\ Cairo University\\
Giza 12613, Egypt\\
Email$^1$: maelzawawy@cu.edu.eg\\
Email$^2$: alanazi@ccis.imamu.edu.sa
 }

\begin{document}

\maketitle                        

\begin{abstract}
Static alias analysis of different type of programming languages has
been drawing researcher attention. However most of the results of
existing techniques for alias analysis are not precise enough
compared to needs of modern compilers. Probabilistic versions of
these results, in which result elements are associated with
occurrence probabilities, are required in optimizations techniques
of modern compilers.

This paper presents a new probabilistic approach for alias analysis
of parallel programs. The treated parallelism model is that of SPMD
where in SPMD, a program is executed using a fixed number of program
threads running on distributed machines on different data. The
analyzed programs are assumed to be in the static single assignment
(SSA) form which is a program representation form facilitating
program analysis. The proposed technique has the form of
simply-structured system of inference rules. This enables using the
system in applications like Proof-Carrying Code (PPC) which is a
general technique for proving the safety characteristics of modern
programs.

\end{abstract}

\vspace{1em}


\section{Introduction}\label{intro}

Considerable efforts of research have been devoted to achieve the
static alias analysis of different type of programming languages.
Algorithms for calculating alias relationships for all program
points exist for basic programming techniques. Classically, alias
relationships fall in two groups: definitely-alias relationships and
possibly-alias relationships. The former is typically true for all
possible execution paths and the later might typically be true for
some of the possible execution paths. However  the  information
calculated by most existing algorithms for alias analysis is not
precise enough compared to the needs of modern compilers. This is so
as modern compilers need finer alias-information to be able to
achieve tasks like code specialization and data speculation. In
other words information calculated by most alias analysis techniques
do not help compilers to do aggressive optimizations. More
specifically, possibly-alias relationships is not rich enough to
inform the comfier about the possibility that constraints for the
executions. Hence compilers are somehow forced to follow a
conservative way and assume the conditions validity for all
execution paths~\cite{KhedkerMR12,El-Zawawy11,ChenHJL04}.

A dominant programming technique of parallelism for large-scale
machines equipped with distributed-memories is the single program,
multiple data (SPMD) model. In SPMD, a program is executed using a
fixed number of program threads running on distributed machines on
different data~\cite{Pacheco11}. SPMD can be executed on
low-overhead and simple dynamic systems and is convenient for
expressing parallelism concepts. This parallelism model is used by
message-sending architectures such as MPI. SPMD is also adapted by
languages whose address spaces are globally partitioned (PGAS) such
as UPC, Co-Array Fortran, and Titanium. Specific deadlocks can be
prevented using the SPDM model which can also be used to achieve
probabilistic data races and specific program
optimizations~\cite{LiFLC13,TsujiSHP13}.

Static single assignment (SSA)~\cite{AmmeHR08,AmmeDFR01} is a
program representation form facilitating program analysis. SSA forms
are important for software re-engineering and compiler construction.
Program analysis needs data-flow information about points of the
program being analyzed. Such information is necessary for program
compilation and re-engineering and is conveniently collected by SSA.
For program variables, some analyses need to know assignment
statements that could have assigned the used variable content. In
Static single assignment (SSA) form exactly one variable definition
corresponds to a variable use. This is only possible if the
algorithm building the SSA form is allowed to insert auxiliary
definitions if it is possible for different definitions to get into
a specific program point.

A general technique for proving the safety characteristics of modern
programs is Proof-Carrying Code
(PCC)~\cite{PfenningCT11,JobredeauxHNF12}. PCC proofs are needed and
typically constructed using logics annotated with inference rules
that are language-specific. The proofs ensures safety in case there
are no bugs in the inference rules. One type of Proof-Carrying Code
is Foundational Proof-Carrying Code (FPCC) which uses theories of
mathematical logic. The small trusted base of FPCCs and the fact
that they are not tied to any specific systems make them more secure
and robust.

\begin{figure*}
\centering  \fbox{
\begin{minipage}{11cm}
{\footnotesize{
\begin{eqnarray*}
& &x\in \hbox{lVar},\ {i_{op}}\in I_{op},\ {b_{op}}\in{ B_{op}},\
\hbox{and }
m\in M\subseteq \mathcal{M}\\
l \in \hbox{Loc}   &::= & {x}\mid {l\rightarrow y}\mid [l] .\\
e \in \hbox{DExpr}   &::= &  {l}\mid  {e_1\ i_{op}\ e_2} \mid {\&l}
\mid {\hbox{malloc}()} \mid {\hbox{run }
(e,m)}\mid \\
& & {\hbox{reform} (\hbox{alis } m , \hbox{int }m)\ e}\mid
{\hbox{reform } (\hbox{int }m_j,
\hbox{int }m_i))\ e}.\\
S \in \hbox{Stmts}   &::= & {l:= e}\mid {\hbox{run } (S,m)} \mid
{S_1;S_2}\mid {x_i:= f(x_j,x_k)} \mid
{x_i:= md(x_j)}\mid \\
& & {mu(x_j)} \mid {\hbox{if }e \hbox{ then } S_t \hbox{ else } S_f}
\mid {\hbox{while } e\ \hbox{do } S_t}.
\end{eqnarray*}
}} \caption{Programming Language Model;
\textit{SSA-DisLang}}\label{f1}
\end{minipage}}\\
\fbox{
\begin{minipage}{11cm}
{\footnotesize{
\[
\begin{prooftree}
P(x)=\{(a_1,p_1),\dots,(a_n,p_n)\}\qquad
i=\max(p_1,\dots,p_n)\justifies x: P\rightarrow_l a_i
\thickness=0.08em\using{(\hbox{x}^p)}
\end{prooftree}
\]\\
\[
\begin{prooftree}
\begin{tabular}{l}
$P(l)=\{(a_1,p_1),\dots,(a_n,p_n)\}$   \qquad
$P(y)=\{(b_1,q_1),\dots,(b_m,q_m)\}$   \\
 $i=\max\{p_j\times q_j\mid a_j\in P(y)\rceil_1\}$
  \end{tabular} \justifies (l\rightarrow y): P\rightarrow_l
b_i \thickness=0.08em\using{(\rightarrow^p)}
\end{prooftree}
\]\\
\[
\begin{prooftree}
\begin{tabular}{l}
$P(l)=\{(a_1,p_1),\dots,(a_n,p_n)\}$   \qquad
$forall i.P(a_i)=\{(b_1^i,q_1^i),\dots,(b_m^i,q_m^i)\}$   \\
 $i=\max\{p_i\times q_j^i\mid 1\le i\le n \& 1\le j\le m\}$
  \end{tabular} \justifies [l]: P\rightarrow_l
b_i \thickness=0.08em\using{([l]^p)}
\end{prooftree}
\]
}} \caption{Probabilistic Alias Analysis (PAA):
 Locations.}\label{f2}
\end{minipage}}\\ \fbox{
\begin{minipage}{11cm}
{\footnotesize{
\[
\begin{prooftree}
e:P\rightarrow a_e\qquad \hbox{probability of arriving at this
memory point}\ge p_{t}\justifies
  {\hbox{reform} (\hbox{alis } m \rightarrow \hbox{int }m)\ e}: P\rightarrow_l
a_e \thickness=0.08em\using{(\hbox{reform}_1^p)}
\end{prooftree}
\]\\
\[\begin{prooftree}
\hbox{probability of arriving at this memory point}< p_{t}\justifies
  {\hbox{reform} (\hbox{alis } m \rightarrow \hbox{int }m)\ e}: P\rightarrow_l
\bot \thickness=0.08em\using{(\hbox{reform}_2^p)}
\end{prooftree}
\]\\
\[
\begin{prooftree}
e:P\rightarrow a_e\qquad \hbox{probability of arriving at this
memory point}\ge p_{t}\justifies
  {\hbox{reform } (\hbox{int }m_j\rightarrow
\hbox{int }m_i))\ e}: P\rightarrow_l a_e
\thickness=0.08em\using{(\hbox{reform}_3^p)}
\end{prooftree}
\]\\
\[\begin{prooftree}
\hbox{probability of arriving at this memory point}< p_{t}\justifies
  {\hbox{reform } (\hbox{int }m_j\rightarrow
\hbox{int }m_i))\ e}: P\rightarrow_l \bot
\thickness=0.08em\using{(\hbox{reform}_4^p)}
\end{prooftree}
\]\\
\[
\begin{prooftree}
e:P\rightarrow a_e\qquad b=\hbox{reform}(\_,\hbox{int }m)a_e
\justifies
 \hbox{run }(e,m): P\rightarrow_l b \thickness=0.08em\using{(\hbox{run}_e^p)}
\end{prooftree}
\]
\[
\begin{prooftree}
a_i\hbox{ is a fresh memory location on machine }m_i \justifies
 \hbox{malloc}(): P\rightarrow_l a_i \thickness=0.08em\using{(\hbox{malloc})^p}
\end{prooftree}
\]
\[
\begin{prooftree}
e_1:P\rightarrow a_{e_1} \qquad
 e_2:P\rightarrow a_{e_2} \justifies
 e_1\ i_{op}\ e_2: P\rightarrow_l a_{e_1}+a_{e_2} \thickness=0.08em\using{(+^p)}
\end{prooftree}
\]
}} \caption{Probabilistic Alias Analysis (PAA): Distributed
Expressions.}\label{f3}
\end{minipage}
}
\end{figure*}
This paper presents a new technique for probabilistic alias analysis
of parallel programs. The technique has the form of
simply-strictured system of inference rules. The information
calculated by the proposed technique are precise enough compared to
information needed by modern compilers for compilations,
re-engineering, aggressive-optimization processes. The proposed
technique is designed to work on the common and robust data-flow
representation; SSA forms of parallel programs. The use of inference
systems in the proposed technique makes it straightforward for our
technique to produce justifications needed by Foundational
Proof-Carrying Code (FPCCs). The proofs have the form of inference
rules derivations that are efficiently transferable. The parallelism
model treated in this paper is that of single program, multiple data
(SPMD) in which the same program is executed on different machines
on different sets of data.

\subsection*{Motivation}
The paper is motivated by need for a precise probabilistic alias
analysis for SPMD programs running on a hierarchy of distributed
machines. The required technique is supposed to associate each
analysis result with a correctness proof (in the form of type
derivations) to be used in proof-carrying code applications.

\subsection*{Contributions}
The contribution of the paper is a new approach for probabilistic
alias analysis of SPMD programs running on SSA forms of programs and
producing justifications with analysis results.

\subsection*{Paper Outline}
The outline of this paper is as follows.  Section~\ref{s1} presents
the langauge model, \textit{SSA-DisLang}, of the paper. This section
also presents an informal semantics to the langauge constructs. The
main content of Section~\ref{s1} is the new technique of the
probabilistic alias analysis of SPMD programs. Section~\ref{s5}
concludes the paper and suggests directions for future work.

\section{Probabilistic Alias analysis for SPMD}\label{s1}

This section presents a new technique for probabilistic alias
analysis of parallel programs. The parallelism model used here is
that of SPMD where the same program is executed on distrusted
machines havering different data. However communications between the
distributed machines is allowed in a predefined contexts. For
example a command running on machine $1$ may request machine $3$ to
evaluate a specific expression using data of machine $3$ and to
return the result to machine $1$.

The syntax of the langauge used to present the new probabilistic
alias analysis technique is shown in Figure~\ref{f1}. We call the
langauge mode \textit{SSA-DisLang} for ease of reference. A program
in \textit{SSA-DisLang} consists of a sequence of statements where
statements are of wide diversity. Statements use (distributed)
expressions, \textit{DExpr}. The machines to run
\textit{SSA-DisLang} programs are typically organized in a
hierarchy. The distributed expressions include the following:
\begin{itemize}
    \item $\hbox{malloc}(): $ allocates a dynamic array in memory and
    return its base address.
    \item $\hbox{run }(e,m):$ evaluates the distributed expression
    $e$ on machine $m$ and return the result.
    \item $\hbox{reform} (\hbox{alis } m , \hbox{int }m)\
    e:$ casts the location denoted by the distributed expression $e$
    as an integer rather than a pointer to a memory location on
    machine $m$.
    \item $\hbox{reform } (\hbox{int }m_j, \hbox{int }m_i))\ e:$
    casts the location denoted by the distributed expression $e$
    as a pointer a memory location on machine $m_i$ rather than as a pointer
    to a memory location on machine $m_j$.
\end{itemize}

Our proposed technique assumes that the given program, that is to be
be analyzed for its probabilistic alias competent, has the static
single assignments form. Therefore the input program would contain
annotations (added by any efficient SSA such as
algorithm~\cite{HungCHJL12}). The program annotations will have the
form of new statements added to the original program. Therefore
statements of \textit{SSA-DisLang} include the following:
\begin{itemize}
    \item ${l:= e}: $ this is a classical assignment command. However
    the design of the langauge allows using this command to
    assign a value evaluated at a machine to a location on  a different machine
    of the machines hierarchy.
    \item ${\hbox{run } (S,m)}:$ allows evaluates a specific command $S$
    on a specific machine $m$ regardless of the executing machine.
    This command is necessary when some commands are convenient only
    to run on data of certain machine of the hierarchy. The command
    is also used when security is a concern as $S$ would not
    have access to all machines.
    \item ${x_i:= f(x_j,x_k)}:$ this command is to be added by the supposed
    SSA algorithm and it semantics is that variable $x_i$ were created
    specifically for avoiding multiple assignments to variable $x$.
    The range of this definition if form definition of variable $x_j$
    to that of variable $x_k$.
    \item ${x_i:= md(x_j)}:$
    this is the second sort of annotations \textit{SSA-DisLang}
    programs. The semantics of this statement is that it is highly likely that
    variable $x_i$ is used to define variable $x_j$. Recall that our
    main technique of the paper cares about possibility of
    assignments to occur in percentages; it is not a $0/1$ technique.
    \item ${mu(x_j)}$: this is the third and last sort of annotations \textit{SSA-DisLang}
    programs. The semantics of this command is that variable $x_j$
    is highly likely to be used in the following de-reference command
    of  the program.
\end{itemize}

\begin{figure*}[htbp]
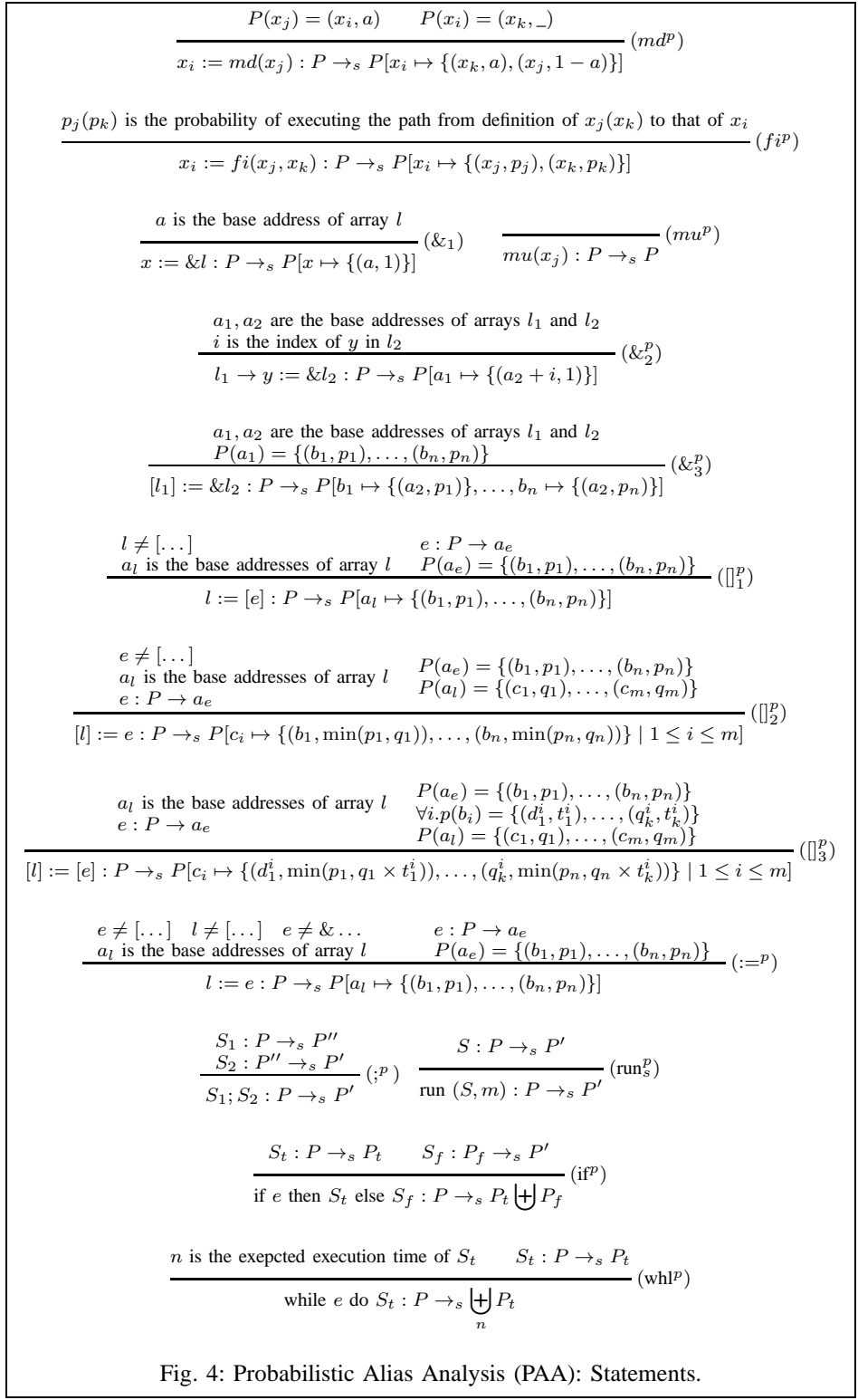

\centering \fbox{
\begin{minipage}{12cm}
{\footnotesize{
\[
\begin{prooftree}
P(x_j)=(x_i,a) \qquad P(x_i)=(x_k,\_) \justifies x_i:= md(x_j):
P\rightarrow_s P[x_i\mapsto\{(x_k,a),(x_j,1-a)\}]
\thickness=0.08em\using{(md^p)}
\end{prooftree}
\]\\
\[
\begin{prooftree}
p_j (p_k) \hbox{ is the probability of executing the path from
definition of  } x_j (x_k) \hbox{ to that of } x_i\justifies
x_i:=fi(x_j,x_k): P\rightarrow_s
P[x_i\mapsto\{(x_j,p_j),(x_k,p_k)\}] \thickness=0.08em\using{(fi^p)}
\end{prooftree}
\]\\
\[
\begin{prooftree}
a \hbox{ is the base address of array }l \justifies x:= \&l:
P\rightarrow_s P[x\mapsto\{(a,1)\}] \thickness=0.08em\using{(\&_1)}
\end{prooftree}
\qquad
\begin{prooftree}
\justifies mu(x_j): P\rightarrow_s P \thickness=0.08em\using{(mu^p)}
\end{prooftree}
\]\\
\[
\begin{prooftree}
\begin{tabular}{l}
 $a_1,a_2 \hbox{ are the base addresses of arrays }l_1 \hbox{ and }l_2$  \\
$i \hbox{ is the index of }y\hbox{ in }l_2$    \\
\end{tabular}
\justifies l_1\rightarrow y := \&l_2: P\rightarrow_s
P[a_1\mapsto\{(a_2+i,1)\}] \thickness=0.08em\using{(\&_2^p)}
\end{prooftree}
\]\\
\[
\begin{prooftree}
\begin{tabular}{l}
$a_1,a_2 \hbox{ are the base addresses of arrays }l_1 \hbox{ and }l_2$  \\
$P(a_1)=\{(b_1,p_1),\dots,(b_n,p_n)\}$    \\
\end{tabular} \justifies [l_1] :=
\&l_2: P\rightarrow_s P[b_1\mapsto
\{(a_2,p_1)\},\dots,b_n\mapsto\{(a_2,p_n)\}]
\thickness=0.08em\using{(\&_3^p)}
\end{prooftree}
\]\\
\[
\begin{prooftree}
\begin{tabular}{l}
$l\not=[\dots]$\\
$a_l\hbox{ is the base addresses of array }l$  \\
\end{tabular}
\begin{tabular}{l}
$e:P\rightarrow a_e$\\
$P(a_e)=\{(b_1,p_1),\dots,(b_n,p_n)\}$    \\
\end{tabular} \justifies l :=
[e]: P\rightarrow_s P[a_l\mapsto \{(b_1,p_1),\dots,(b_n,p_n)\}]
\thickness=0.08em\using{([]_1^p)}
\end{prooftree}
\]\\
\[
\begin{prooftree}
\begin{tabular}{l}
$e\not=[\dots]$\\
$a_l\hbox{ is the base addresses of array }l$  \\
$e:P\rightarrow a_e$\\
\end{tabular}
\begin{tabular}{l}
$P(a_e)=\{(b_1,p_1),\dots,(b_n,p_n)\}$    \\
$P(a_l)=\{(c_1,q_1),\dots,(c_m,q_m)\}$    \\
\end{tabular}
\justifies [l]:= e:P\rightarrow_s P[c_i\mapsto
\{(b_1,\min(p_1,q_1)),\dots,(b_n,\min(p_n,q_n))\}\mid 1\le i\le m]
\thickness=0.08em\using{([]_2^p)}
\end{prooftree}
\]\\
\[
\begin{prooftree}
\begin{tabular}{l}
$a_l\hbox{ is the base addresses of array }l$
\\$e:P\rightarrow a_e$\\
\end{tabular}
\begin{tabular}{l}
$P(a_e)=\{(b_1,p_1),\dots,(b_n,p_n)\}$    \\
$\forall i.p(b_i)=\{(d_1^i,t_1^i),\dots,(q_k^i,t_k^i)\}$\\
$P(a_l)=\{(c_1,q_1),\dots,(c_m,q_m)\}$    \\
\end{tabular}
\justifies [l]:= [e]:P\rightarrow_s P[c_i\mapsto
\{(d_1^i,\min(p_1,q_1\times t_1^i)),\dots,(q_k^i,\min(p_n,q_n\times
t_k^i))\}\mid 1\le i\le m] \thickness=0.08em\using{([]_3^p)}
\end{prooftree}
\]\\
\[
\begin{prooftree}
\begin{tabular}{l}
$e\not=[\dots]$\quad $l\not=[\dots]$\quad$e\not=\&\dots$\\
$a_l\hbox{ is the base addresses of array }l$
\end{tabular}\qquad
\begin{tabular}{l}$e:P\rightarrow a_e$\\
$P(a_e)=\{(b_1,p_1),\dots,(b_n,p_n)\}$    \\
\end{tabular}
\justifies l:= e:P\rightarrow_s P[a_l\mapsto
\{(b_1,p_1),\dots,(b_n,p_n)\}] \thickness=0.08em\using{(:=^p)}
\end{prooftree}
\]\\
\[
\begin{prooftree}
\begin{tabular}{l}
$S_1: P\rightarrow_s P^{\prime\prime} $\\
$S_2: P^{\prime\prime}\rightarrow_s P^\prime $
\end{tabular}
\justifies S_1;S_2: P\rightarrow_s P^\prime
\thickness=0.08em\using{(;^p)}
\end{prooftree}
\quad
\begin{prooftree}
S: P\rightarrow_s P^\prime \justifies \hbox{run } (S,m):
P\rightarrow_s P^\prime \thickness=0.08em\using{(\hbox{run}_s^p)}
\end{prooftree}
\]\\
\[
\begin{prooftree}
S_t: P\rightarrow_s P_t \qquad S_f: P_f\rightarrow_s P^\prime
\justifies \hbox{if }e \hbox{ then } S_t \hbox{ else } S_f:
P\rightarrow_s P_t\biguplus P_f
\thickness=0.08em\using{(\hbox{if}^p)}
\end{prooftree}
\]\\
\[
\begin{prooftree}
n \hbox{ is the exepcted execution time of }S_t\qquad S_t:
P\rightarrow_s P_t \justifies \hbox{while } e\ \hbox{do } S_t:
P\rightarrow_s \biguplus_n
P_t\thickness=0.08em\using{(\hbox{whl}^p)}
\end{prooftree}
\]
}} \caption{Probabilistic Alias Analysis (PAA):
Statements.}\label{f5}
\end{minipage}
}
\end{figure*}

Figures~\ref{f2},~\ref{f3}, and~\ref{f5} present elements of our
proposed technique for probabilistic alias analysis of
\textit{SSA-DisLang} programs. The proposed technique has the form
of a type system which consists of set of alias types denoted by $P$
and set of inference rules presented in the technique figures. An
alias type is a \textit{partial} map. The domain of this partial map
is a subset of the set of all variables (denoting registers) allowed
to be used on different machines of the distributed hierarchy plus
the set of all addresses of memories of machines on hierarchy.  The
codomain of the alias type is the power set of the set of all
\textit{probabilistic pairs}. A probabilistic pair is a pair of
variable (register) or a memory location and a number $p$ such hat
$0\le p\le 1$.

Judgment produced by the system have the forms $e:P\rightarrow a$
and $S:P\rightarrow P^\prime$. The judgement $e:P\rightarrow a$
means that evaluating the expression $e$ in a memory state of the
type $P$ results in the memory address $a$. The semantics of the
judgement $e:P\rightarrow a$ is that running $S$ in a memory state
of the type $P$ results (if ends) in a memory state of the type
$P^\prime$. The proposed technique is meant to be used as follows.
given a distributed program $S$, one constructs (using inference
rules of the system) an alias type $P^\prime$ such that
$S:\bot\rightarrow P^\prime$. The base type is the partial map with
an empty domain is denoted by $\bot$. The construction of $P^\prime$
is a type derivation process and results in annotating program with
the required probabilistic alias information.

Inference rules for distributed expressions are shown in
figure~\ref{f3}. Some comments on the rules are in order. The rules
for \textit{reform} expressions only considers the address evaluated
from $e$ if there is a considerable probability (probability
threshold $>p_{th}$) of arriving at the concerned program point.

Inference rules for statements are shown in figure~\ref{f5}. Some
comments on the rules are in order. The rule $([]^p_3)$ uses the
base address of the array denoted by $l$ and the address returned
for $e$ by the inference rules of expressions. The image of these
addresses under the pre-type also contribute to calculating the post
type of the de-reference statement.

The soundness of our proposed technique is guarantied by the
following theorem. The theorem requests the existence of robust
operational semantics for the langauge \textit{SSA-DisLang}. Many
semantics candidates exist. Due to lack of space we only reference
to the semantics in this paper. From the authors's experience and
based on some experiments, the simplicity of the theorem proof
deeply relies on the choice of the langauge semantics.
\begin{theorem}
Suppose that $S$ is a \textit{SSA-DisLang} program and
$S:\bot\rightarrow P^\prime$. Suppose also that using a convenient
operational semantics for \textit{SSA-DisLang}, the execution of $S$
is captured as $S:M\rightarrow M^\prime$. Then the final memory
state $M^\prime$ is of the the probabilistic alias type $P^\prime$.
\end{theorem}

\section{Related Work}\label{s4}

The changing associations characteristics property of pointers makes
the points-to analysis a complicated problem~\cite{ChenHJL04}. Much
research~\cite{Ben-AsherR13,Staiger-Stohr13,LiCK13,HuangLW13} have
been developed to solve the pointer analysis problem. Each of these
techniques evaluates either points-to or aliases relationships at
program points. Points-to  and aliases relationships are classified
into two classes: definitely-aliases (or must-points-to)
relationships and may-points-to (or possibly-aliases relationships).
While the later relationships are true on some executions, the
former relationships are true on all executions. Wether
possibly-aliases or may-points-to relationships are true on most
executions or on few executions is not measurable by most of these
techniques. For specific transformations and optimizations these
missed information are beneficial. Few attempts were made to fill
this gap.

Using traditional data-flow analysis,
in~\cite{ShaoCZ08,MiwakeichiVABWMY07} a theoretical formulation is
presented to compute measurable information. More specifically, this
work evaluates, for each program point, the predicted count that
specific conditions may hold. Aiming at evaluating, among array
references, the probabilities of
aliases,~\cite{JuCO99,GuoWWBOVCA06,FangCOW06} presents a
probabilistic technique for memory disambiguation. A probabilistic,
interprocedural, contextsensitive, and flow-sensitive techniques for
alias analysis were proposed
in~\cite{ChenHJL04,HungCHJL12,PierroHW07}. On alias relationships,
these technique evaluate measurable information. MachSUIF and SUIF
compiler infrastructures provided the bases for the implementation
of these techniques. The probabilities of pointer induced, loop
carried, and data dependence relationships were evaluated
in~\cite{ChenHHJL03,ZhaiSCM08}. Using sparse matrices, as efficient
linear transfer functions,~\cite{SilvaS06,RoyS10} modeled
probabilistic alias analysis. The results of this research were
proved accurate. ~\cite{LuC11} presents an algorithm to evaluate
measurable alias information. A technique for memory disambiguation,
evolution of probabilities that pairs of memory pointers point at
the same memory location, is presented in~\cite{LuC11}.

For array optimizations and analysis, probabilistic techniques for
memory disambiguation were proposed~\cite{JuCO99}. These techniques
typically present data speculations~\cite{XiangS13} necessary for
modern architectures of computers.

For distributed parallel machines with shared-memory, an important
problem is that of compiler optimizations for programs that are
pointer-based. This is so as the host processor of an object can be
determined using data distribution analysis~\cite{LeeHC97} and
affinity analysis~\cite{CarlisleR96}.

In, pointer-based programs, a reference is referencing a group of
objects with may-points-to. For such cases, traditional affinity
analysis~\cite{Pitkaranta09} can be integrated with traditional
pointer analysis. The result of this integration is a technique that
evaluates the parts of objects on a processor's list of task
executions. This is necessary for many program optimizations.

There are many examples of aggressive optimizations such as data
speculations, speculative multithreading (thread partitioning), and
code specialization~\cite{ZhangDS12,Khan10}. To boost performance of
modern architectures, these optimizations are typically achieved by
compilers. Compilers can only do such tasks if they are able to
measure the possibility of dynamic pointer associations. Using
interval analysis, irreducible flow graphs, and the elimination
technique, intraprocedural analysis can be used to handle pointer
analysis of programs~\cite{SunZC11}. Extensions to such techniques
to cover context-sensitive analysis that is interprocedural is
achievable as well.

Examples of analysis for speculative multithreading model include
thread partitioning~\cite{LiuZLSF14,LiZYD11,KimCC11}. Such analysis
boosts compilers performance via running speculative threads in case
of low possibilities for conflicts. In this scenario for threads
with high possibilities are turned off~\cite{ChenHHJL03}.

\section{Conclusion and Future Work}\label{s5}
This paper presented a new technique for probabilistic alias
analysis of SPMD programs. The new approach has the form of system
of inference rules. This has direct applications in proof-carrying
code area of research. The proposed technique also has the advantage
of assuming SSA forms of analyzed programs.

Directions for future work include the following. Producing
probabilistic techniques for important analyses (such as dead-code
elimination) for SPMD programs that uses the results of the analysis
proposed in this paper would be an important contribution. Producing
other analyses for the langauge model of this paper in the spirit of
~\cite{El-Zawawy12-6,El-Zawawy12-5,El-Zawawy11-3} is another
direction for future work. There is also a need for precise
probabilistic operational semantics for SPMD programs. This
semantics would be important to accurately measure probabilities of
statements executions and probabilities of executions order.

\section{Acknowledgment}
The authors acknowledge the support (grants numbers 340918 \&
330911) of the deanship of scientific research of Al Imam Mohammad
Ibn Saud Islamic University (IMSIU).


\bibliographystyle{IEEEtran}
\bibliography{Xbib}

\end{document}